\documentclass[12pt]{article}%
\usepackage{amsmath,latexsym}
\usepackage{graphicx}
\usepackage{amsmath}
\usepackage{amsfonts}
\usepackage{amssymb}%
\begin{document}

\title{{Slowly evolving noncommutative-geometry
        wormholes}}
   \author{
Peter K.F. Kuhfittig*\\  \footnote{kuhfitti@msoe.edu}
 \small Department of Mathematics, Milwaukee School of
Engineering,\\
\small Milwaukee, Wisconsin 53202-3109, USA}

\date{}
 \maketitle

\begin{abstract}
This paper discusses noncommutative-geometry
wormholes in the context of a cosmological model
due to Sung-Won Kim.  An ansatz suggested by the
Friedmann-Lemaitre-Robertson-Walker (FLRW) model
leads to the assumption that the matter content
can be divided into two parts, a cosmological
part depending only on time and a wormhole part
depending only on space.  These assumptions are
sufficient for deriving a complete zero-tidal
force wormhole solution.  The wormhole is
evolving due to the scale factor in the FLRW
model; it is restricted, however, to the
curvature parameters $k=0$ and $k=-1$.  Unlike
previous models, the noncommutative-geometry
background affects both the wormhole part and
the cosmological part of the solution.

\end{abstract}

\section{Introduction}\label{E:introduction}

Wormholes are handles or tunnels in spacetime
connecting widely separated regions of our
Universe or different universes altogether.
While there had been some forerunners,
macroscopic traversable wormholes were first
discussed in detail by Morris and Thorne
\cite{MT88} in 1988.  A few years later,
Sung-Won Kim \cite{Kim} proposed the
possible existence of an evolving wormhole
in the context of the
Friedmann-Lemaitre-Robertson-Walker (FLRW)
cosmological model by assuming that the
matter content can be divided into two
parts, the cosmological part that depends on
time only and the wormhole part that depends
on space only.  The discussion was later
expanded by Cataldo et al. \cite{mC13}.

The purpose of this paper is to study the
relationship between wormholes inspired by
noncommutative geometry and the Kim model.
The noncommutative-geometry background
simultaneously affects both the wormhole
construction and the cosmological part of
the solution.  This result differs
significantly from the outcomes in Refs.
\cite{Kim} and \cite{mC13}.

Regarding the strategy, Ref. \cite{MT88}
concentrates mainly on the wormhole
geometry by specifying the metric
coefficients.  This strategy requires a
search for matter or fields that can produce
the energy-momentum tensor needed to sustain
the wormhole.  Here it needs to be emphasized
that we are able to satisfy the geometric
requirements from the physical properties.
The result is an evolving zero-tidal force
wormhole solution; it is restricted to the
curvature parameters $k=0$ and $k=-1$,
corresponding to an open Universe.

Viewed from a broader perspective, it has
already been shown that noncommutative
geometry, which is an offshoot of string
theory, can account for the flat galactic
rotation curves \cite{pK17, fR12}, but
under certain conditions, noncommutative
geometry can also support traversable
wormholes \cite{KG14, pK18, pK16, pK15,
 Jamil14, FKRI12}.

This paper is organized as follows: Sec.
\ref{S:structure} briefly recalls the
structure of wormholes and the basic
features of noncommutative geometry.
Sec. \ref{S:Kim} continues with the
Sung-Won Kim model. Here the discussion
is necessarily more detailed, partly
in the interest of completeness, but mainly
to allow the inclusion of a more general
form of the Einstein field equations.
These are subsequently used in Sec. \ref
{S:special} to obtain a wormhole solution
that does not depend on the separation
of the matter content.  In Sec. \ref{S:nc}
we derive a wormhole solution from the
noncommutative-geometry background,
followed by a discussion of the null
energy condition in Sec. \ref {S:violation}.
Sec. \ref{S:comparison} features a
 comparison to an earlier solution.
 In Sec. \ref{S:conclusion}, we conclude.
%END OF SECTION

\section{Wormhole structure and
   noncommutative geometry}\label{S:structure}

Morris and Thorne \cite{MT88} proposed the
following static and spherically symmetric
line element for a wormhole spacetime:
\begin{equation}\label{E:line1}
ds^{2}=-e^{2\Phi(r)}dt^{2}+\frac{dr^2}{1-b(r)/r}
+r^{2}(d\theta^{2}+\text{sin}^{2}\theta\,
d\phi^{2}),
\end{equation}
using units in which $c=G=1$.  Here $b=b(r)$
is called the \emph{shape function} and
$\Phi=\Phi(r)$ is called the \emph{redshift
function}, which must be everywhere finite
to avoid an event horizon.  For the shape
function we must have $b(r_0)=r_0$, where
$r=r_0$ is the radius of the \emph{throat}
of the wormhole.  The wormhole spacetime
should be asymptotically flat, i.e.,
$\text{lim}_{r\rightarrow \infty}\Phi(r)
=0$ and  $\text{lim}_{r\rightarrow \infty}
b(r)/r=0$.  An important requirement is the
\emph{flare-out condition} at the throat:
$b'(r_0)<1$, while $b(r)<r$ near the
throat.  The flare-out condition can only
be met by violating the null energy
condition (NEC), which states that
\begin{equation}
  T_{\alpha\beta}k^{\alpha}k^{\beta}\ge 0
\end{equation}
for all null vectors $k^{\alpha}$, where
$T_{\alpha\beta}$ is the energy-momentum
tensor.  Matter that violates the NEC is
called ``exotic" in Ref. \cite{MT88}.  In
particular, for the outgoing null vector
$(1,1,0,0)$, the violation has the form
\begin{equation}
   T_{\alpha\beta}k^{\alpha}k^{\beta}=
   \rho +P^r<0.
\end{equation}
Here $T^t_{\phantom{tt}t}=-\rho$ is the energy
density, $T^r_{\phantom{rr}r}= P^r$ is the
radial pressure, and
$T^\theta_{\phantom{\theta\theta}\theta}=
T^\phi_{\phantom{\phi\phi}\phi}=P^t$ is
the lateral pressure.

Returning now to the noncommutative-geometry
background mentioned earlier, we need to
recall that, as an offshoot of string theory,
noncommutative geometry replaces point-like
particles by smeared objects.  (For a detailed
discussion, see Refs. \cite {SS03, NSS06,
NS10}.)  As a result, spacetime can be
encoded in the commutator
$[\textbf{x}^{\mu},\textbf{x}^{\nu}]
=i\theta^{\mu\nu}$, where $\theta^{\mu\nu}$ is
an antisymmetric matrix that determines the
fundamental cell discretization of spacetime
in the same way that Planck's constant
discretizes phase space \cite{NSS06}.  An
interesting and effective way to model the
smearing effect, discussed in Refs.
\cite{pK15,LL12, NM08}, is to assume that
the energy density of the static,
spherically symmetric, smeared, and
particle-like gravitational source is
given by
\begin{equation}\label{E:rho}
  \rho(r)=\frac{\mu\sqrt{\beta}}
     {\pi^2(r^2+\beta)^2},
\end{equation}
which can be interpreted to mean that the
gravitational source causes the mass $\mu$
of a particle to be diffused throughout the
region of linear dimension $\sqrt{\beta}$
due to the uncertainty; so $\sqrt{\beta}$
has units of length.
(Ref. \cite{NSS06} uses a Gaussian
distribution instead of Eq. (\ref{E:rho})
to represent $\rho$.)  Eq. (\ref{E:rho})
leads to the mass distribution
\begin{equation}\label{E:source}
   \int^r_04\pi(r')^2\rho(r')dr'=
   \frac{2M\sqrt{\beta}}{\pi}
   \left(\frac{1}{\sqrt{\beta}}\text{tan}^{-1}
   \frac{r}{\sqrt{\beta}}-
   \frac{r}{r^2+\beta}\right),
\end{equation}
where $M$ is now the total mass of the source.

According to Ref. \cite{NSS06}, noncommutative
geometry is one of the basic properties of
spacetime that does not depend on particular
features such as curvature.  Moreover, since
the noncommutative effects can be implemented
by modifying only the energy-momentum tensor,
there is no need to change the Einstein tensor
in the field equations.  As a result, the
length scales need not be microscopic.
%END OF SECTION

\section{The Sung-Won Kim model}\label{S:Kim}

The Sung-Won Kim cosmological model with a
traversable wormhole is given by \cite{Kim}
\begin{equation}\label{E:line2}
ds^{2}=-e^{2\Phi(r)}dt^{2}+[R(t)]^2\left[
\frac{dr^2}{1-kr^2-b(r)/r}
+r^{2}(d\theta^{2}+\text{sin}^{2}\theta\,
d\phi^{2})\right],
\end{equation}
where $R(t)$ is the scale factor of the
Universe and $k$ is the sign of the
curvature of spacetime, i.e., $k=+1$,
0, or $-1$.   (So while $b=b(r)$ is still
the shape function, it is subject to
conditions that are different from those
of a Morris-Thorne wormhole.)  The
Einstein field equations in Ref. \cite
{Kim} are based on Eq. (\ref{E:line2}),
but it is subsequently assumed that
$\Phi(r)\equiv 0$ to be consistent with
the FLRW model.

The discussion of the Sung-Won Kim model
is continued and elaborated on in Ref.
\cite{mC13}.  Unfortunately, the field
equations in Refs. \cite{Kim} and
\cite{mC13} do not agree and need to be
rederived.  Here we follow Ref. \cite{Kim}
and base the calculations on line element
(\ref{E:line2}).  The results are
\begin{equation}\label{E:Einstein1}
   8\pi\rho(r,t)=3\left(\frac{\dot{R}}{R}\right)^2
   e^{-2\Phi(r)}+\frac{3k}{R^2}+\frac{b'(r)}{R^2r^2},
\end{equation}
\begin{equation}\label{E:Einstein2}
   8\pi P^r(r,t)=-2\frac{\ddot{R}}{R}e^{-2\Phi(r)}
   -\left(\frac{\dot{R}}{R}\right)^2e^{-2\Phi(r)}
   -\frac{k}{R^2}-\frac{b(r)}{R^2r^3}+\frac{2}{R^2r}
   \Phi'(r)\left(1-kr^2-\frac{b(r)}{r}\right),
\end{equation}
\begin{multline}\label{E:Einstein3}
   8\pi P^t(r,t)=-2\frac{\ddot{R}}{R}e^{-2\Phi(r)}
   -\left(\frac{\dot{R}}{R}\right)^2e^{-2\Phi(r)}
   -\frac{k}{R^2}+\frac{b(r)-rb'(r)}{2R^2r^3}\\
   +\frac{1}{R^2}\left[(\Phi'(r))^2\left(1-kr^2-
   \frac{b(r)}{r}\right)+\Phi''(r)\left(1-kr^2-
   \frac{b(r)}{r}\right)\right.\\
   \left.-\frac{1}{2}\Phi'(r)\left(2kr
   +\frac{rb'(r)-b(r)}{r^2}\right)\right]
   +\frac{1}{R^2r}\Phi'(r)\left(1-kr^2
   -\frac{b(r)}{r}\right),
\end{multline}
\begin{equation}\label{E:Einstein4}
  8\pi T_{01}=\frac{\dot{R}}{R^2}
  e^{-\Phi(r)}\Phi'(r)
  \left(1-kr^2-\frac{b(r)}{r}\right)^{1/2},
\end{equation}
where $T_{01}$ is the outward energy flow.
(The prime and overdots denote the derivatives
with respect to $r$ and $t$, respectively.)

If $\Phi(r)\equiv 0$, the results agree with
those in Ref. \cite{mC13} (omitting $\Lambda$,
the cosmological constant).  For convenience,
these will now be restated:
\begin{equation}\label{E:E1}
   8\pi\rho(r,t)=3\left(\frac{\dot{R}}{R}\right)^2
   +\frac{3k}{R^2}+\frac{b'}{R^2r^2},
\end{equation}
\begin{equation}\label{E:E2}
   8\pi P^r(r,t)=-2\frac{\ddot{R}}{R}-
   \left(\frac{\dot{R}}{R}\right)^2
   -\frac{k}{R^2}-\frac{b}{R^2r^3},
\end{equation}
\begin{equation}\label{E:E3}
   8\pi P^t(r,t)=-2\frac{\ddot{R}}{R}
   -\left(\frac{\dot{R}}{R}\right)^2
   -\frac{k}{R^2}+\frac{b-rb'}
   {2R^2r^3}.
\end{equation}

The next step depends on a key insight,
due to Sung-Won Kim \cite{Kim}, that
allows the separation of the Einstein
field equations into two parts, namely,
the following ansatz for the matter
parts:
\begin{equation}\label{E:Kim1}
   R^2(t)\rho(r,t)=R^2(t)\rho_c(t)
       +\rho_w(r),
\end{equation}
\begin{equation}\label{E:Kim2}
   R^2(t)P^r(r,t)=R^2(t)P_c(t)
      +P^r_w(r),
\end{equation}
\begin{equation}\label{E:Kim3}
   R^2(t)P^t(r,t)=R^2(t)P_c(t)
      +P^t_w(r).
\end{equation}
The subscripts $c$ and $w$ refer,
respectively, to the cosmological
and wormhole parts.  So $P_c$
necessarily represents the isotropic
pressure.

The ansatz now allows us to separate
Eqs. (\ref{E:E1})-(\ref{E:E3}) into
two parts, the left side being a
function of $t$ and the right side
a function of $r$, also carried out in
Ref. \cite{mC13}:
\begin{equation}\label{E:separate1}
   R^2\left[8\pi\rho_c-3\left(\frac{\dot{R}}
   {R}\right)^2-\frac{3k}{R^2}\right]=
   \frac{b'}{r^2}-8\pi\rho_w=l,
\end{equation}
\begin{equation}\label{E:separate2}
  R^2\left[8\pi P_c+2\frac{\ddot{R}}{R}
  +\left(\frac{\dot{R}}{R}\right)^2
  +\frac{k}{R^2}\right]=-\frac{b}{r^3}
  -8\pi P^r_w=m,
\end{equation}
\begin{equation}\label{E:separate3}
   R^2\left[8\pi P_c+2\frac{\ddot{R}}{R}
   +\left(\frac{\dot{R}}{R}\right)^2
   +\frac{k}{R^2}\right]=
   \frac{b-rb'}{2r^3}-8\pi P^t_w=m,
\end{equation}
where $l$ and $m$ are constants.  The
reason is that a function of $t$ cannot
be equal to a function of $r$ for all
$t$ and $r$ unless they are equal to
some constant.  In Eqs. (\ref{E:separate2})
and (\ref{E:separate3}), the constants
are the same since the cosmological parts
are equal.  The constants $l$ and $m$ may
be taken as arbitrary.

Returning now to Eqs.
(\ref{E:Einstein1})-(\ref{E:Einstein4}),
recall that we obtained the more general
form of the field equations by using
line element (\ref{E:line2}), as
suggested in Ref. \cite{Kim}.  It now
becomes apparent, however, that the
separation in Eqs.
(\ref{E:separate1})-(\ref{E:separate3})
cannot be carried out by means of Eqs.
(\ref{E:Kim1})-(\ref{E:Kim3}) unless
we assume that $\Phi(r)\equiv 0$, which
takes us back to the FLRW model.
%END OF SECTION

\section{The noncommutative wormhole}
   \label{S:nc}
To obtain a wormhole solution, we will
consider the special case $l=-3m$,
following Ref. \cite{mC13}.  Eqs.
(\ref{E:separate1})-(\ref{E:separate3})
then yield
\begin{equation}\label{E:R1}
  3\left(\frac{\dot{R}}{R}\right)^2
   +\frac{3k}{R^2}-\frac{3m}{R^2}
   =8\pi\rho_c
\end{equation}
and
\begin{equation}\label{E:R2}
   -2\frac{\ddot{R}}{R}-
   \left(\frac{\dot{R}}{R}\right)^2
   -\frac{k}{R^2}+\frac{m}{R^2}
   =8\pi P_c
\end{equation}
for the cosmological part, while the
wormhole part is given by
\begin{equation}\label{E:W1}
   \frac{b'}{r^2}-8\pi\rho_w=-3m,
\end{equation}
\begin{equation}\label{E:W2}
   -
\frac{b}{r^3}-8\pi P^r_w=m,
\end{equation}
\begin{equation}\label{E:W3}
   \frac{b-rb'}{2r^3}-8\pi P^t_w=m.
\end{equation}

The wormhole solution can be obtained
from Eqs. (\ref{E:W1})-(\ref{E:W3}) by
making use of Eq. (\ref{E:rho}).  It
remains to be seen what restrictions
will be placed on the solutions due to
the cosmological part and to the
necessary violation of the NEC.  But
for now we have from Eq. (\ref{E:W1})
and Eq. (\ref{E:rho}) that
\begin{equation}
    b'(r)=8\pi\frac{\mu\sqrt{\beta}r^2}
    {\pi^2(r^2+\beta)^2}-3mr^2.
\end{equation}
Integrating and using the condition $b(r_0)
=r_0$, we obtain
\begin{multline}\label{E:shape2}
  b(r)=\frac{4M\sqrt{\beta}}{\pi}
  \left(\frac{1}{\sqrt{\beta}}\text{tan}^{-1}
  \frac{r}{\sqrt{\beta}}-\frac{r}{r^2+\beta}
  \right)-mr^3\\
  -\frac{4M\sqrt{\beta}}{\pi}
  \left(\frac{1}{\sqrt{\beta}}\text{tan}^{-1}
  \frac{r_0}{\sqrt{\beta}}-\frac{r_0}{r_0^2
  +\beta}\right)+mr_0^3+r_0,
\end{multline}
where $M$ is the mass of the wormhole.
It now becomes apparent that the resulting
spacetime is not asymptotically flat.  The
normal procedure is to cut off the wormhole
material at some $r=a$ and then join the
structure to an external Schwarzschild
spacetime. We will see in the next section,
however, that the need to violate the NEC
($\rho +P^r<0$) requires a slight
modification of the shape function,
resulting in the required asymptotic
flatness.
%END OF SECTION

\section{Violating the NEC}\label{S:violation}
To check the violation of the NEC
($\rho +P^r<0$) for the wormhole, we
let $R\equiv \text{constant}$
and obtain from Eqs. (\ref{E:E1}) and
(\ref{E:E2}),
\begin{equation}
   b-rb'-2kr^3>0.
\end{equation}
At $r=r_0$, we therefore get
\begin{equation}
   r_0-r_0\frac{8\pi\mu\sqrt{\beta}r_0^2}
   {\pi^2(r_0^2+\beta)^2}+3mr_0^3
   -2kr_0^3>0.
\end{equation}
Since $\sqrt{\beta}$ is extremely small, we
actually have
\begin{equation}\label{E:NEC}
   r_0+(3m-2k)r_0^3\gtrsim 0.
\end{equation}
To check this condition, we need to return
to Ref. \cite{NSS06} for some additional
observations.  The relationship between
the radial pressure and energy density
is given by
\begin{equation}\label{E:EoS}
   P^r=-\rho.
\end{equation}
The reason is that the source is a
self-gravitating droplet of anisotropic
fluid of density $\rho$ and the radial
pressure is needed to prevent the
collapse back to the matter point.  In
addition, the  lateral pressure is
given by
\begin{equation}\label{E:tr1}
   P^t=-\rho-\frac{r}{2}
     \frac{\partial\rho}{\partial r}.
\end{equation}
Since the length scales can be
macroscopic, we can retain Eq.
(\ref{E:EoS}) and then use Eq.
(\ref{E:tr1}) to write
\begin{equation}\label{E:tr2}
    P^t=-\rho-\frac{r}{2}
     \frac{\partial\rho}{\partial r}
     =P^r+\frac{2\mu r^2\sqrt{\beta}}
     {\pi^2(r^2+\beta)^3}
\end{equation}
by Eq. (\ref{E:rho}).  So on larger
scales, we have $P^r=P^t$.  Since the
pressure becomes isotropic, we can
assume the equation of state to be
$P_c=-\rho_c$.  Substituting in Eqs.
(\ref{E:R1}) and (\ref{E:R2}), we get
\begin{equation*}
   -2\frac{\ddot{R}}{R}
   +2\left(\frac{\dot{R}}{R}\right)^2
   +\frac{2k}{R^2}-\frac{2m}{R^2}=0.
\end{equation*}
This equation can be rewritten as
\begin{equation}
   3\frac{\ddot{R}}{R}
   -3\left(\frac{\dot{R}}{R}\right)^2=
   \frac{3k}{R^2}-\frac{3m}{R^2}.
\end{equation}
Subtracting the Friedmann equations
\[
   3\frac{\ddot{R}}{R}=-4\pi(\rho_c
      +3P_c)
\]
and
\[
   3\left(\frac{\dot{R}}{R}\right)^2
   =8\pi\rho_c-\frac{3k}{R^2}
\]
now yields
\begin{equation}
   \frac{3k}{R^2}-\frac{3m}{R^2}=
   -4\pi(\rho_c+3P_c)-8\pi\rho_c
   +\frac{3k}{R^2}.
\end{equation}
So if $P_c=-\rho_c$, we obtain
\begin{equation}
   m=0, \text{independently of}\quad k.
\end{equation}
Applied to Eq. (\ref{E:NEC}), the NEC
is violated if
\begin{equation}
   k=0\quad \text{or}\quad k=-1.
\end{equation}
These conditions correspond to an open
Universe.

To summarize, we employed basic physical
principles to derive the following
zero-tidal force solution:
\begin{equation}
   \Phi(r)\equiv 0
\end{equation}
and (since $m=0$)
\begin{multline}\label{E:shape1}
  b(r)=\frac{4M\sqrt{\beta}}{\pi}
  \left(\frac{1}{\sqrt{\beta}}\text{tan}^{-1}
  \frac{r}{\sqrt{\beta}}-\frac{r}{r^2+\beta}
  \right)\\
  -\frac{4M\sqrt{\beta}}{\pi}
  \left(\frac{1}{\sqrt{\beta}}\text{tan}^{-1}
  \frac{r_0}{\sqrt{\beta}}-\frac{r_0}{r_0^2
  +\beta}\right)+r_0.
\end{multline}
The slowly evolving wormhole solution is
restricted to the values $k=0$ and $k=-1$
to ensure that the NEC is violated.  The
wormhole spacetime is asymptotically flat.

\section{The special case $k=0$}
    \label{S:special}
For completeness let us briefly consider
a wormhole solution that does not depend
on the separation of the Einstein field
equations.  We can combine Eqs.
(\ref{E:E1}) and (\ref{E:E2})
to obtain
\begin{equation}
   8\pi r^3R^2\left[\rho(r,t)+P^r(r,t)
   \right]=2r^3(\dot{R}^2-R\ddot{R})
   +2r^3k+rb'(r)-b(r).
\end{equation}
If we now let $k=0$, then Eq.
(\ref{E:line2}) represents an evolving
Morris-Thorne wormhole with the usual
shape function $b=b(r)$.  The NEC is
violated at the throat $r=r_0$ for all
$t$ whenever
\begin{equation}\label{E:NEC1}
   8\pi r_0^3R^2\left[\rho(r_0,t)-
   P^r(r_0,t)\right]=2r_0^3(\dot{R^2}
   -R\ddot{R})+r_0b'(r_0)-b(r_0)<0.
\end{equation}
If the Universe is indeed accelerating,
then the term $-R\ddot{R}$ eventually
becomes dominant due to the
ever-increasing $R$.  So for
sufficiently large $R$, the NEC is
violated, thereby fulfilling a key
requirement for the existence of
wormholes.  (Inflating Lorentzian
wormholes are discussed in Ref.
\cite{tR93}.)

Recalling that the radial tension $\tau$
is the negative of $P^r$, Inequality
(\ref{E:NEC1}) can be written (since
$b(r_0)=r_0$)
\begin{equation}\label{E:NEC2}
   8\pi r_0^2R^2\left[\tau(r_0)-\rho(r_0)
   \right]=2r_0^2(-\dot{R^2}+R\ddot{R})
   -b'(r_0)+1>0.
\end{equation}
If $R(t)\equiv 1$, this reduces to the
static Morris-Thorne wormhole; so if
$b'(r_0)<1$, then $\tau(r_0)>\rho(r_0)$,
requiring exotic matter.  In Inequality
(\ref{E:NEC2}), however, $\tau(r_0)>
\rho(r_0)$ could result from the
dominant term $R\ddot{R}$.  In that
case, the NEC is violated without
requiring exotic matter  for the
construction of the wormhole itself.
%END OF SECTION

\section{Comparison to an earlier solution}
    \label{S:comparison}
A wormhole solution inspired by
noncommutative geometry had already been
considered in Ref. \cite{pK15}.  The
Einstein field equation $\rho(r)=
b'(r)/(8\pi r^2)$, together with Eq.
(\ref{E:rho}), leads directly to the
static solution, Eq. (\ref{E:shape1}).
(Here it is understood that $k=0$,
but $R(t)$ could be retained.)
Unfortunately, this simple approach
leaves the redshift function
undetermined.  The desirability of
zero tidal forces then suggested the
assumption $\Phi(r)\equiv 0$ in Ref.
\cite{pK15}.  It is shown in Ref.
\cite{pK09}, however, that this
assumption causes a Morris-Thorne
wormhole to be incompatible with the
Ford-Roman constraints from quantum
field theory.  Given the
noncommutative-geometry background,
rather than the purely classical setting
in Ref. \cite{MT88}, this objection
does not apply directly.

It is interesting to note that in the
present paper, the zero-tidal force
solution is built into the Sung-Won
Kim model and does not require any
additional considerations.
%END OF SECTION

\section{Conclusion}\label{S:conclusion}
Morris-Thorne wormholes typically require
a reverse strategy for their theoretical
construction: specify the geometric
requirements and then manufacture or search
the Universe for matter or fields to obtain
the required energy-momentum tensor.  One
of the goals in this paper is to obtain a
complete wormhole solution from certain
physical principles.  To this end, we assume
a noncommutative-geometry background, as in
previous studies, but we also depend on a
cosmological model due to Sung-Won Kim that
is based on the FLRW model with a
traversable wormhole.  The basic assumption
is that the matter content can be divided
into two parts, a cosmological part that
depends only on $t$ and a wormhole part
that depends only on the radial coordinate
$r$.  The result is a complete zero-tidal
force solution; it is restricted, however,
to the values $k=0$ and $k=-1$, corresponding
to an open Universe.  This conclusion is
consistent with the special case $k=0$
discussed in Sections \ref{S:special}
and \ref{S:comparison}.

The wormhole is slowly evolving due to the
scale factor $R(t)$ and, critically, the
noncommutative-geometry background not
only produces the wormhole solution, it
also affects in a direct manner the
cosmological part of the solution.  This
conclusion differs significantly from
those in Refs. \cite {Kim} and \cite{mC13}.

\end{document}